# Study of charging-up of PCB planes for neutrino experiment readout

B. Baibussinov,[a] M. Bettini,[a] F. Fabris,[a] A. Guglielmi,[a] S. Marchini,[a] G. Meng,[a] M. Nicoletto,[a] F. Pietropaolo, [a,b] G. Rampazzo,[a] R. Triozzi,[a] and F. Varanini[a,1]

[a]*Sezione INFN di Padova e Università di Padova,*
  *via Marzolo 8, 35131 Padova, Italy*
[b]*CERN,*
  *Route de Meyrin, 1211 Geneva, Switzerland*
  *E-mail*: filippo.varanini@pd.infn.it

ABSTRACT: The use of double-faced, metallized, perforated PCB planes, segmented into strips for the anodic read-out of ionization signals in liquid argon TPCs, is emerging as a promising technology for charge readout in liquid argon TPCs used in large volume detectors.As a proof of concept, a prototype liquid Argon TPC hosting this new anode configuration based on single side perforated PCB planes has been constructed and exposed to cosmic rays at LNL in Italy. Tests were performed with both the metallized and insulating sides of the anode facing the drift volume, providing the first evidence of the focusing effect on drift electron trajectories through the PCB holes due to charge accumulation on the insulator surface.

---

[1] Corresponding author.

# Contents

## 1. Introduction

The Liquid Argon Time Projection Chamber (LAr-TPC) detection technique, introduced by C. Rubbia [1], is extensively used in particle physics to study rare events and neutrino interactions. These detectors provide high granularity imaging by collecting with mm pitch anodic wire chambers ionization electrons, produced by charged particles in LAr and drifted by a uniform electric field. A 3D imaging of ionizing events is achieved through stereoscopic and non-destructive read-out of the TPC anodes, which are composed of a sequence of parallel planes (typically three) with wires oriented in different directions. Drifting electrons crossing the first planes produce an "induced" bipolar currents on the sensing wires before being collected in the last "collection" plane. The geometry and voltage biases of the planes ensure the transparency to drifting electrons in all crossed planes and the full charge collection on the last one. The ICARUS T-600 760 t detector took data in LNGS underground laboratory in 2010-13, exposed to the CNGS neutrino beam [2], and is presently in operation at Fermilab with Booster and NuMI beams [3].

The LAr-TPC technology has been evolving continuously over the last few decades in preparation for constructing multikton detectors for the next generation of underground astroparticle and long baseline neutrino oscillation experiments. The use of double-faced, metallized, perforated PCB planes segmented into strips for the anodic read-out of ionization signals in liquid argon TPCs (*Charge Readout Plane, CRP*) represents a promising alternative to the wire-based chambers offering advantages such as better mechanical stability and lower cost per channel.

A first prototype of CRP was initially developed and successfully tested at INFN Legnaro National Laboratory (LNL) [4]. Subsequent intensive R&D activities, focusing on scaling up of the CRP concept led to the design of the future DUNE Vertical Drift LAr-TPC, which exploits the CRP anode system [5]. Similar to wires, this anode read-out system provides 3D reconstruction of a large variety of ionizing events through stereoscopic and non-destructive TPC signal read-out, based on a sequence of three planes with strips oriented in different directions.

This paper focuses on a further evolution of the setup consisting of PCB planes metalized and divided into strips on only one face. In this configuration, charge carriers (either electrons or argon ions) resulting from the ionization of the liquid argon are expected to be partially collected on the non-metalized face of the planes, while electrons are read out on the conductive metalized side. This charge deposition is expected to accumulate over time, modifying the electric field near the PCB planes. As a result the focusing of the drifting electrons trajectories through the PCB holes and the transparency of the crossed anodic planes are further enhanced.

## 2. Description of the chamber

The LARISA (Liquid Argon Integrated Solid Anode) readout system consists of three identical perforated PCB planes in a regular hexagonal shape (**Figure 1**) with copper read-out strips running at $0, \pm 60^0$ orientations etched on one side of PCB support for optimal 3D reconstruction of ionization events. The ionization electrons drift across the first two planes and are collected on the last one. The hexagonal pattern is suitable for scaling up to large read-out anodic surfaces required in multi-kton LAr-TPCs.

The features of the detector signals depend on the electric field in the anodic region which is determined by the hole radius **r**, the distance **d** between adjacent holes, the thickness **t** of the PCB layer, the spacing **s** between successive PCB planes, and the voltage applied to the various read-out planes. The parameters of LARISA set-up, chosen on the basis of a good transparency, technical feasibility and cost, are summarized in **Table 1**. A drawing of a plane and a detail of the strips can be seen in **Figure 1.** The strips are connected to the amplifier through a 1 nF x 3kV capacitor. Each strip of a layer are connected to their common bias through a 10 MΩ resistor.

| PCB thickness (**t**) | 1.6 mm |
|---|---|
| Spacing between successive PCBs (**s**) | 5 mm (between closest sides) |
| Hole radius (**r**) | 2.1 mm |
| Distance between adjacent holes (**d**) | 6 mm (in each direction) |
| Holes to total surface ratio (**R**) | 0.44 |
| Strip width (**w**) | 6 mm |
| Strip orientation wrt vertical | 0º, 60º, -60º (I1, I2, C) |

**Table 1.** Summary of the geometrical parameters selected for chamber construction.

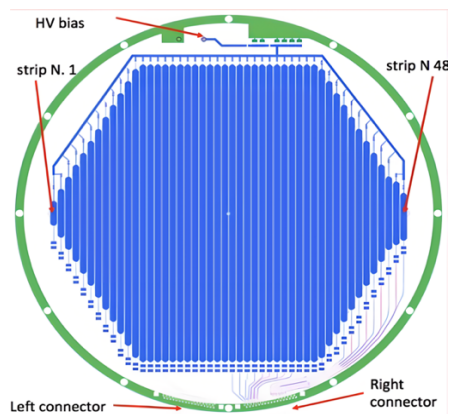

**Figure 1.** Scheme of the metalized strips on a face of a PCB plane (the holes in the strips are not shown here). The progressive numbering of the strips, their connection to the bias HV and to the connectors for signal read-out are indicated.

The three identical PCB planes are mounted together to compose the full read-out system of the chamber, with a spacing of 5 mm between planes and with the strips orientation as illustrated in Figure **2**. A detail of the strips and holes is shown in **Figure 3**. The Induction 1 layer, facing the TPC drift volume, is followed by Induction 2 and Collection layers.

Two "standard" and "inverted" orientations of the chamber with respect to the electron drift direction, illustrated in **Figure 4**, were tested. In the first one, the metallized side of each PCB was facing the drift volume while in the "inverted" configuration the insulated PCB side faced the drift volume.

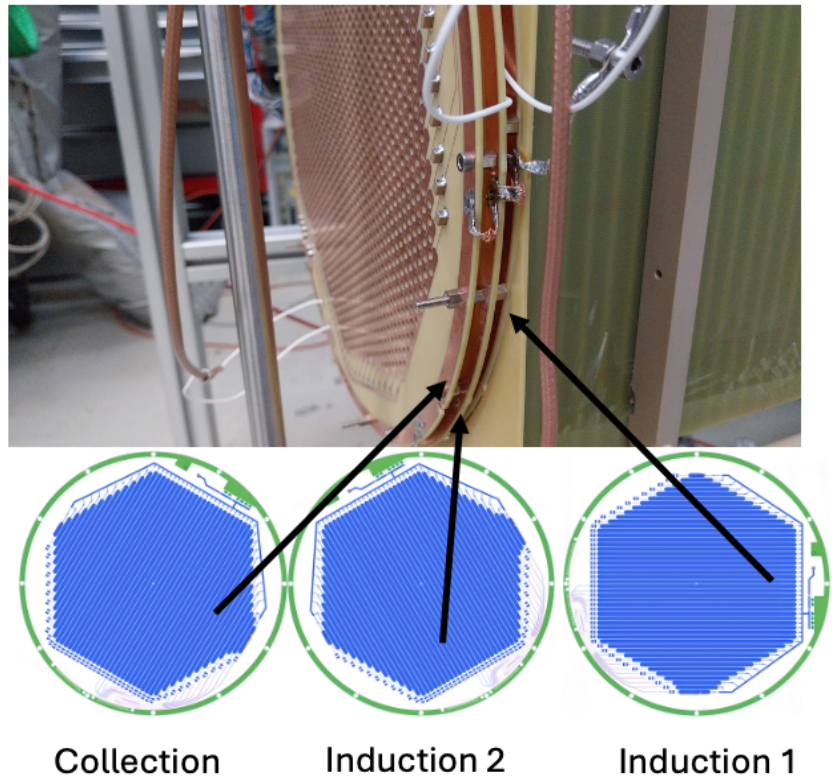

**Figure 2.** Photo of the fully assembled LARISA Chamber. The drift volume (a cube of ~30 cm side) is visible on the right, while the anode read-out system is on the left. The edges of the 3 PCB planes can be identified, with the metalized side oriented away from the drift volume as shown in the bottom left picture. The flat connectors carrying signals from the strips and the cables carrying the voltage bias can also be observed.

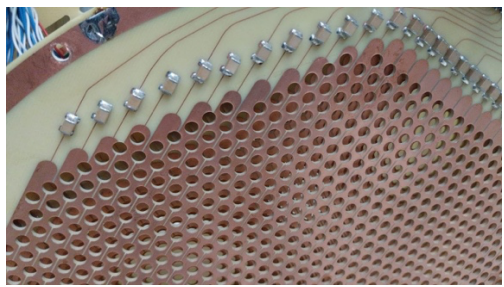

**Figure 3.** Detail of one of the PCB planes. The strips and holes are visible, as the tracks connecting the strips to read-out electronics.

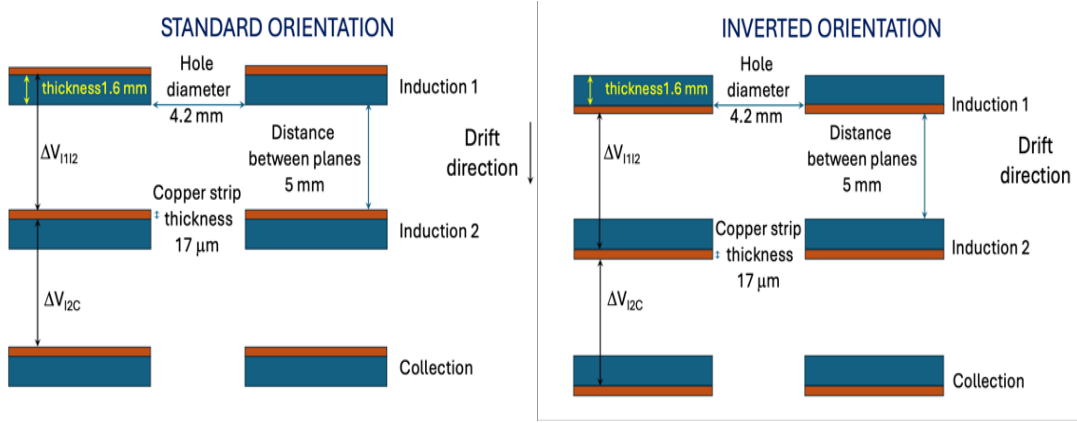

**Figure 4.** Schematic of the "standard" (left) and "inverted" (right) orientations of the anode read-out system.

## 3. Simulations

A simulation of the drift of ionization electrons in liquid Argon under an external electric field $E_{DRIFT}$ and their collection on the electrodes was performed using the COMSOL Multiphysics$^{TM}$ package for the detector geometry described in **Section 2**.

A complete 3D map of the electric field in the liquid argon volume near the read-out electrodes was computed. Ionization electrons were generated in the drift volume, and their trajectories were traced along the electric field until they were collected on the last plane or on an upstream Induction plane. The transparency of the Induction planes was evaluated as the fraction of drift electrons crossing the planes relative to the total generated electrons.

In the LARISA set-up all the PCB perforations are optically aligned along the same axis (**Figure 4**). In the absence of electrostatic focusing, the transparency of a read-out plane is determined by the fraction R~ 0.44 of the electrode surface corresponding to holes over the total surface (see Table 1). Transparency can be enhanced if the electron trajectories are focused into the holes by an electric field between the anode plains exceeding $E_{DRIFT}$ as illustrated in **Figure 5**. Due to optical alignment of the holes, the electron trajectories after focusing through the first Induction plane, cross the successive ones unperturbed. Transparency has been studied as a function of single voltage differences $\Delta V_{PCB}$ applied between the first and second Induction planes and between the second Induction and the Collection planes.

In both standard and inverted configurations, two limiting cases were considered regarding the charge accumulation on the non-metalized sides. The first case refers to the initial condition of the chamber, just after the bias voltage on the planes is turned on, with no charge present on the insulating surfaces. The second case describes the "asymptotic charging-up" conditions, where charge accumulation has reached a stable equilibrium preventing further accumulation due to electrostatic repulsion from the charge already deposited on the PCB surface. The electrostatic field near the anode evolves from the no charging-up condition to the asymptotic one over a timescale depending on the rate of any ionization source in the active volume (cosmic rays, radioactive source, …).

The transparency of the LARISA anode set-up as a function of $\Delta V_{PCB}$ is shown in **Figure 6,** for both standard and inverted orientations, under no charging-up and asymptotic charging-up conditions.

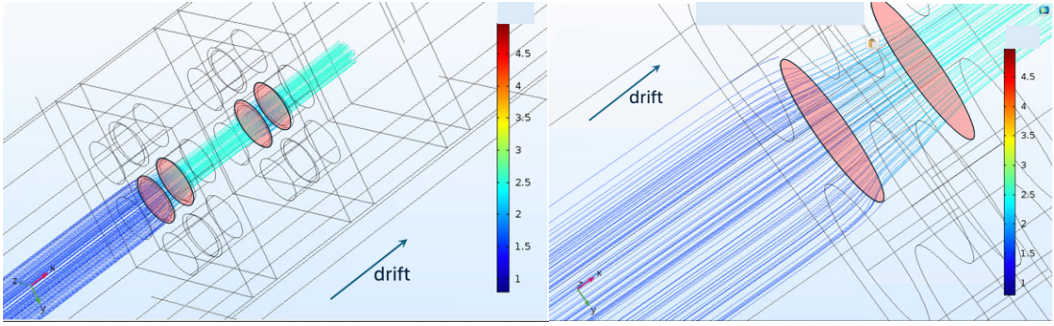

**Figure 5.** Illustrations of the trajectories of 100 drift electrons in a PCB scheme with a standard configuration (described in the text), assuming ΔV=1100 V and no charging-up. The transparency of the first induction plane is 88% in this case while the one of Induction 2 is full (see Figure 5). The color scale refers to the drift velocity (in mm/μs). In the zoomed image (on the right), some lines ending on the Induction 1 plane, resulting in inefficiencies, can be seen. In both images, the upstream and downstream ends of the holes that are crossed by the drift electrons, separated by the PCB thickness, are marked in red. Holes in the collection plane are not shown for simplicity.

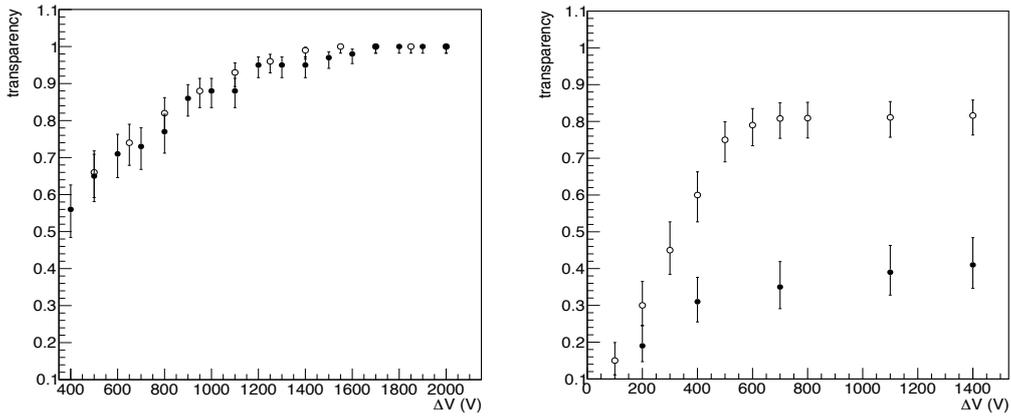

**Figure 6.** Left: Transparency of the Induction 1 plane obtained in simulations for the standard chamber orientation, in no charging-up conditions (full dots) and asymptotic charging-up conditions (empty dots). Right: Transparency of the Induction 1 plane obtained in simulations for the inverted chamber orientation, in no charging-up conditions (full dots) and asymptotic charging-up conditions (empty dots).

The impact of charging-up is quite limited for the standard configuration, resulting in only a few percent increase in transparency, whereas it is very significant for the inverted configuration. This outcome is expected since the charging-up of the non-metalized PCB side facing the drift volume helps improving the funneling of the electron trajectories into the PCB holes. In the standard orientation, the accumulated charge is shielded by the metallic layers facing the drift, so the focusing is only due to the metal. Note that full transparency which is not reached in the inverted configuration due to the 1.6 mm thickness of the PCB, can be achieved by reducing this thickness to 0.5 mm.

## 4. Experimental set-up and data-taking campaign

The LARISA PCB anodic structure was installed inside an existing general-purpose test facility for liquid Argon TPCs (ICARINO) located at the INFN-LNL laboratories in Italy which has been previously used for various tests of TPC read-out technologies, electronics and cryogenics (more details in [7]).

ICARINO consists of a liquid Argon TPC with an active volume of ~27 liters characterized by ~ 30 cm drift distance and square field cage with 30 cm sides (see **Figure 7**). It is equipped with a HV system to generate uniform drift electric field up to 500 V/cm and a bias system for the read-out electrodes, as well as cabling to external analog and digital electronics for signal read-out.

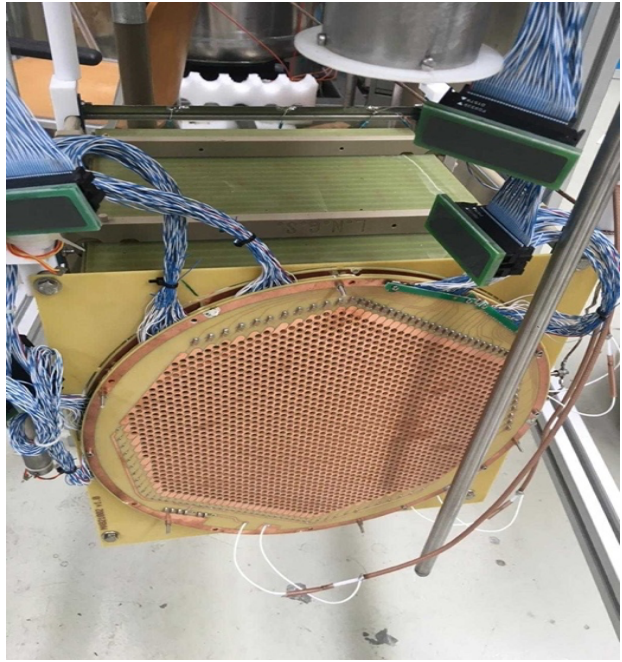

**Figure 7.** Photo of the LARISA 3-plane anodic readout system mounted on the TPC. The Collection plane is visible in the foreground, as the cables carrying the signal from strips and the bias on each plane. The field cage can be seen in the background.

A system of external plastic scintillation counters equipped with photomultipliers provides trigger signals selecting almost vertical cosmic rays crossing the TPC at rate of ~0.04 Hz. A photo of the experimental set-up is in **Figure 8**.

The read-out electronics uses the boards developed by CAEN for the ICARUS-T600 run at LNGS. The V791C version of the analog front-end boards characterized by a signal shaping function with a peaking time of ~3 μs has been used, followed by the V789 board for signal digitization. For this electronic set-up, the gain is ~5.5 ADC counts per fC; a more detailed description can be found in [2].

The liquid recirculation and purification system, based on an Oxisorb filter, reduces the concentration of electronegative impurities in the Argon volume that can attenuate the ionization signal, quantified by a drifting electron lifetime $\tau_{ele}$. During the operations this parameter is measured regularly observing the charge attenuation along cosmic muon tracks crossing the drift volume, using the algorithm described in [7]. The resulting value of $\tau_{ele}$ is used to correct the deposited energy depending on the position along the drift coordinate.

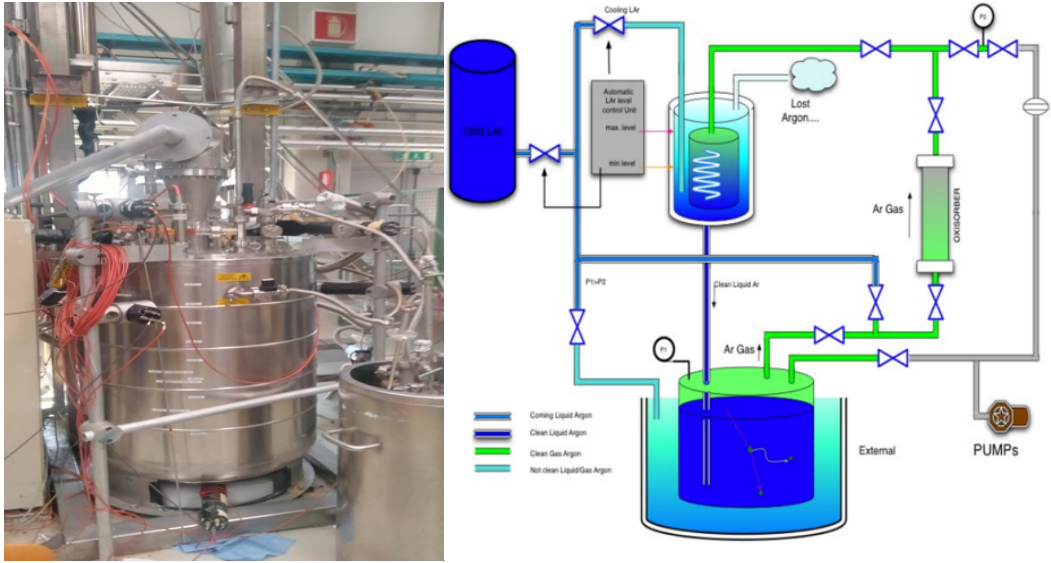

**Figure 8.** Left: a photo of the ICARINO set-up used for both LARISA data-taking campaigns. The cryostat containing the TPC is visible at the center. The 5 PMTs used for the trigger (4 on the sides and 1 under the cryostat) can also be seen, as the purification cartridges on the top of the picture. Right: a scheme of the ICARINO cryogenic and Argon purification plant.

Two LARISA data taking campaigns have been performed, Run 1 for about a month with the "standard orientation", and Run 2 for ~15 days with the "inverted orientation". Since the electron lifetime did not vary significantly during Run 1 and Run 2, a single value of $\tau_{ele}$ for each of these periods (388±94 and 465±60 µs respectively) was used to apply the purity correction.

## 5. Study of induction signals as a function of ΔV (Run 1)

Through-going cosmic muon tracks were recorded to quantify the transparency of the perforated anode structure, the bipolar nature of Induction signals and the charge collection signal on the last plane. A study of the PCB structure transparency was performed increasing the voltage difference $\Delta V_{PCB}$ from 650 V to 2000 V, in steps of 150 V. A muon track collected with $\Delta V_{PCB}$ = 1550 V is shown in **Figure 9**.

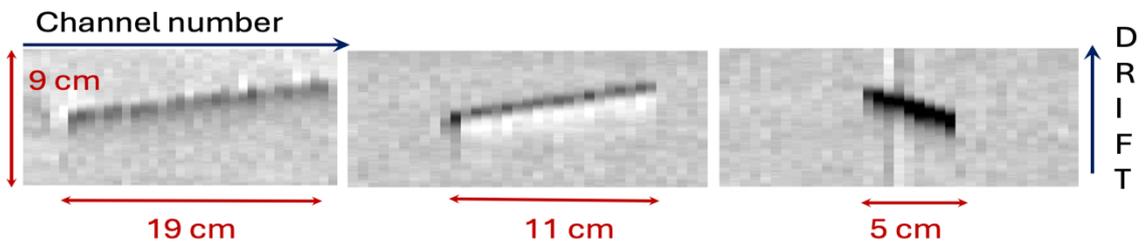

**Figure 9.** Event display of a muon track visualized on all three PCB planes (left to right: Induction 1, Induction 2, Collection) for ΔV=1550 V. Each pixel represents a strip (6 mm) on the horizontal coordinate and a time sample from the electronics (400 ns ~ 0.6 mm) on the vertical one. In the Collection and Induction 2 planes, black indicates a positive signal and white a negative one. In the induction 1 picture, the color scale is inverted (black=negative, white=positive) to ensure a better visibility.

The typical signals from strips in each view are reported in **Figure 10**. The intensity of the trailing lobe of the Induction 1 signal indicates of the fraction of electrons focused into the holes. In the absense of efficient focusing, the trailing lobe is less intense than expected, and a "collection-like" signal component appears from the electrons hitting the Induction 1 plane (see **Figure 11**).

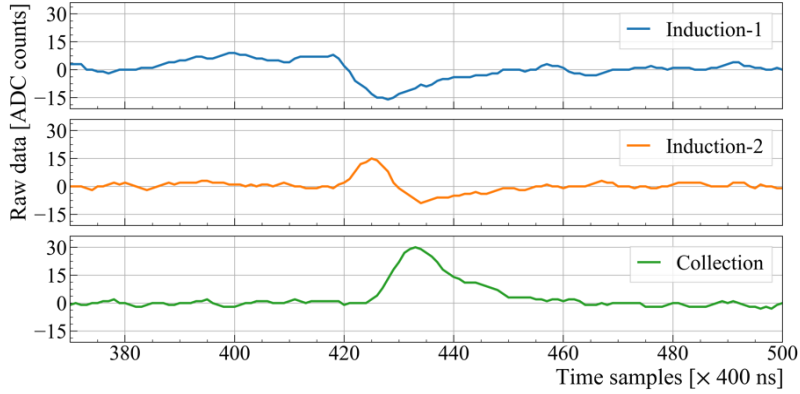

**Figure 10.** Waveforms corresponding to the signals on the first strips in each view (at the top in the image) from the muon in **Figure 9** (top to bottom: Induction 1, Induction 2, Collection).

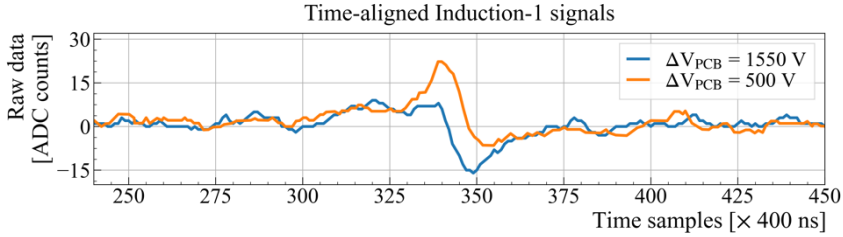

**Figure 11.** Signal from a muon track acquired with ΔV=500 V on one Induction 1 strip (in orange) compared with the signal from a run with high ΔV and good transparency (the same as in Figure 10, in blue). The positive signal corresponds to the collection of the drifting electrons on the PCB, while the negative one corresponds to a signal produced by electrical induction by crossing electrons.

Passing straight muon tracks were selected requiring the geometrical alignment of the hits reconstructed in Induction 1 (quantified by a linear correlation coefficient larger than 0.8). The signals recorded on all three planes allowed for full 3D spatial track reconstruction. The energy deposition was estimated from the measurement of Collection signal areas converted into energy using the calibration constant of the readout electronics as determined in [2].

Electron recombination in liquid Argon was accounted for by introducing a 0.69 correction factor typical of MIP muons, which is an average value for cosmic muons at sea level at 500 V/cm (see [8]), while the effect of the attenuation due to electronegative impurities was corrected as described in the previous section.

The direct estimation of the transparency of Induction planes is given by the ratio $R_{CHG}$ of the charge reaching the Collection plane over the calculated value in the case of full transparency. The ratio between the most probable value of the dE/dx Landau distribution from cosmic muons recorded in the LARISA chamber and the expected one for cosmics[1] is used to evaluate $R_{CHG}$. The values measured in Run 1 as a function of the applied voltage $\Delta V$ are shown in **Figure 12** compared with the transparency expected from simulation. A full transparency is reached at ~1600 V, in fair agreement with predictions from simulation (**Figure 6**). The small difference between the asymptotic charging-up and no charging-up cases cannot be disentangled within the present experimental resolution.

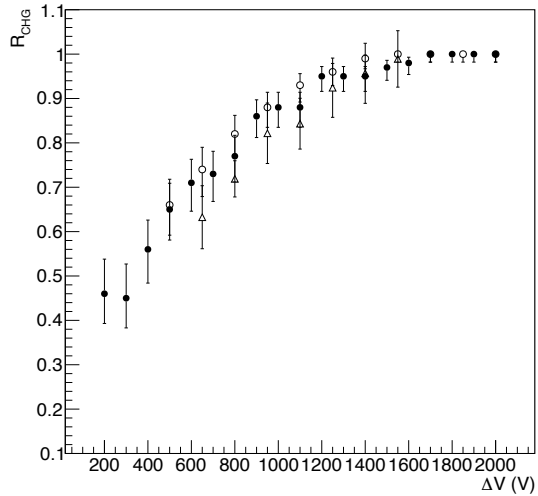

**Figure 12.** $R_{CHG}$ values observed in data for Run 1 (empty triangles), compared to transparency values computed in simulations for both no charging up and asymptotic charging up conditions (full and empty dots respectively).

## 6. Charging-up study with inverted configuration

A second data-taking campaign was conducted under the same conditions as Run 1, but with the chamber in the inverted configuration (see **Section 2**). Data for this study were collected starting ~2 days after turning on the chamber HV.

The transparency of the induction planes, measured using the ratio $R_{CHG}$ defined in **Section 5**, showed a trend as a function of $\Delta V$ that was in full agreement with asymptotic charging-up conditions (see **Figure 13**). This result indicates that charge accumulation on the non-metalized PCB side of the Induction 1 plane had quickly reached asymptotic charging-up conditions, most lickely in less than one hour due to exposure to cosmic rays on surface. This result represent the first observation of the transparency enhancement from charging-up on insulating surface in liquid Argon. A transparency plateau is reached at $\Delta V$~600 V (see **Figure 13**) which is much smaller than the $\Delta V$~1600 V of the non-inverted set-up. The transparency at the plateau is not complete, due to the large thickness of the PCB plane, but it could be improved in future tests with a thinner PCB.

The test demonstrated that the LARISA anodic read-out could be operated stably, allowing non-destructive read-out at voltages as low as a few hundreds of volts, as charging-up corrects the focusing inefficiency.

---

[1] For the expected MPV, a value of 1.83 MeV/cm was obtained from literature for the average direction of the muons in LARISA, with a ~1% uncertainty due to the wide cosmic muon energyspectrum, a negligible contribution respect to other experimental errors.

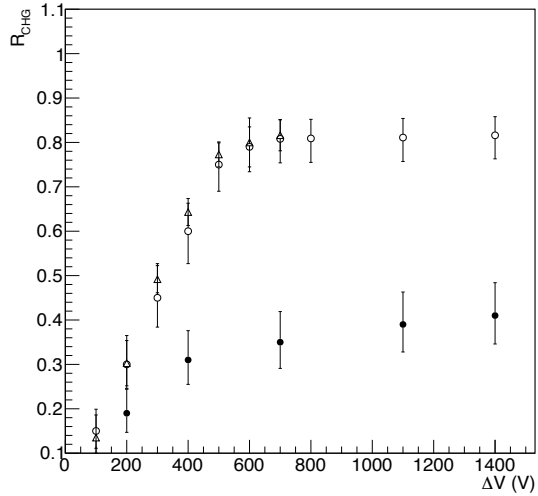

**Figure 13.** R<sub>CHG</sub> value observed in data for Run 2 (empty triangles), compared to transparency values computed in simulations for both no charging up and asymptotic charging up conditions (full, empty dots respectively).

## 7. Conclusions and perspectives

Cosmic ray data collected with a LAr-TPC equipped with the LARISA anodic read-out system, which is based on three perforated PCB layers with copper strips running at different orientations, can provide non-destructive signals allowing for 3D reconstruction of events with millimeter-scale resolution, similarly to traditional wire chambers.

In the standard orientation of metallized PCB side, full transparency is achieved with 1.6 kV voltage difference across the planes, as expected from the hole size and density. In this configuration, charging-up effects on the insulating surface, do not significantly impact the transparency of the upstream Induction plane.

In the inverted PCB orientation, an 80% transparency plateau is reached at a much lower voltage (0.6 kV) because of the charging-up of the insulating PCB side facing the drift volume.

Full transparency is not obtained due to the 1.6 mm thickness of the PCB, whereas it is expected for a 0.5 mm thick plane. However it has been demonstrated that charging-up significantly increases the transparency of Induction planes. This effect can be exploited to recover inefficiencies and allow this read-out technology to operate at a lower bias voltage.

New tests with the LARISA setup are planned to better characterize these features by using thinner PCB planes. Additionally, cryogenic read-out electronics based on a input stage could be deployed to provide a cleaner signal waveform [7].

## Acknowledgements

The substantial support of INFN to the LARISA chamber study and realization in term of starting grant to one of the authors (F. Varanini) is acknowledged. The technical support by the Electronics Service of the INFN Padova Unit, that realized the design of the PCB layout, including the geometry of the strip connections and of the voltage bias, is recognized.